# Space-based weather observatory at Earth-Moon Lagrange point L1 to monitor earth's magnetotail effects on the Moon

*Saurabh Gore [1], Manuel Ntumba [2]*

*[1][2] Division of Space Applications, Tod'Aérs - Transparent Organization for Aeronautics and Space Research, Lomé, Région maritime, Togo.*

*[1] Department of Aerospace Engineering, Moscow Aviation Institute (National Research University), Moscow, Russian Federation.*

*[2] Space Generation Advisory Council, Vienna, Austria.*

sdgore@mai.education

manuel.ntumba@spacegeneration.org

## ABSTRACT

Lunar hematite is formed by the oxidation of iron on the surface of the Moon by oxygen from the Earth's upper atmosphere. The Moon's surface is continuously affected by solar particles from the sun. However, Earth's magnetic tail blocks 99 % of the solar wind and provides windows of opportunity to transport oxygen from Earth's upper atmosphere to the Moon through magnetotail when it is in its full moon phase. Here, we propose to place a space weather observatory at the Earth-Moon L1 Lagrange point carrying a crucial payload onboard to study how Earth's magnetotail causes the Moon's surface to rust. The space weather observatory monitors the effect of Earth's magnetic field on the Moon using advanced spectroscopic sensors from Lagrange-based stations. Earth-moon L1 Lagrange point is the key location for space-weather observation as spacecraft near this point obtains a nearly unobstructed view of the moon. Numerical methods needed for a high-order analytical approximation have been implemented for more accurate predictions.

## 1. Introduction

Chandrayaan-1 lunar mission launched by ISRO in November 2008 led to many discoveries including the discovery of hematite at higher lunar latitudes (> 60 °) [15] and the impact of Earth's magnetotail on the Moon. [2][3][4][5][6][7] The Moon Mineralogy Mapper (m3) onboard Chandrayaan-1 collected reflectance data between 0.46 and 2.98 µm at a spectral resolution of 20 to 40 nm in global mapping mode. [14] This allowed us to search for hematite that confirmed the oxidation process taking place at the lunar surface. The mission objective was to produce a complete map of the surface chemical composition and three-dimensional topography with Polar Regions as regions of interest because they may contain water ice. [8][9][10][11][12] More water trace is observed at higher latitudes which may cause oxidization of the iron present in lunar soil. However, Upstream solar wind (mostly contains H) directly impacts the surface of the moon due to the lack of atmosphere which acts as a reducer to the oxidation process. [13] A recent study shows that a sufficient amount of oxygen may have been delivered from the earth's upper atmosphere to the lunar surface during the passage of the moon through the earth's magnetotail. [18] The magnetosphere is a region surrounding a space object in which charged particles are affected by that object's magnetic field and created by its active inner dynamic. [21][22] Earth's magnetic field lines are affected by solar winds, the flow of electrically conductive plasma, as emitted by the Sun. [23][24] To block the solar and cosmic radiations, Earth releases charged particles opposing the flow of the solar wind, which leads to elongated magnetotail well beyond the orbit of the moon. [25] Earth-Moon L1 Lagrange point is an excellent location for monitoring the magnetotail effects of the Earth's magnetic field on the Moon. The L1 point of the Earth-moon system affords an uninterrupted view of the moon.

## 2. Mission Concept

As Earth's magnetotail extends beyond the orbit of the moon, the moon passes through magnetotail three days before it is full. It takes about 5-6 days to cross the magnetotail. During the pass, the moon comes in contact with charged particles with oxygen and other gasses trapped in the tail from the upper atmosphere of the earth. [14] Measurement of flow of these charged particles and their chemical composition can be done from the onboard instruments of the spacecraft placed in this magnetospheres' region. Earth-moon Lagrange point L1 is the ideal location to carry out this weather observation due to its position in the area of interest as spacecraft lies approximately in line of alignment of Sun-Earth-Moon when the transportation of the oxygen takes place through the upstream solar wind to the surface of the moon. [14][18]

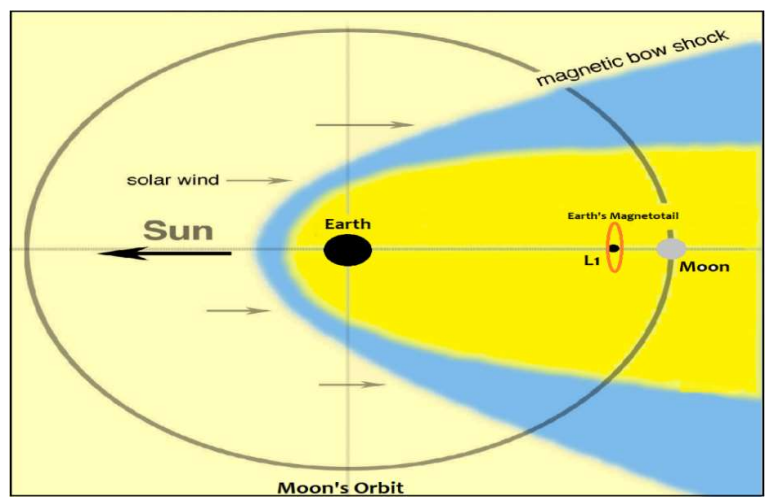

**Figure 1. Space weather observation from L1 langrage point during moon's passage through magnetotail. (Image Credit: NASA)**

Space weather observatory orbiting at Lagrange point L1 equipped with crucial payload and Instruments will lead to a deep understanding of the geologic, topographic, and mineralogical features of the lunar surface. Charged Particle Spectrometers (CPS) onboard the spacecraft could help with measuring the irradiation of Solar and Galactic Cosmic Ray, elemental composition, energy characteristics of gamma rays produced by the lunar surface. Terrain Imager, Multi-band Imager, and Spectral Profiler can continuously observe the moon and provide critical stereoscopic observations. Surface composition study with help of stereo images taken from the terrain imager will help us to reveal geologic information of lunar surface. The environment of magnetic field and upstream solar wind from earth to the moon and around the moon can be studied by the lunar magnetometer, Electron Spectrum Analyzer, Ion mass, and energy analyzer. An ultraviolet telescope could provide data of the Earth's upper atmosphere and Plasmasphere from the orbit which helps us to detect resonantly scattered emissions from oxygen and helium ion.[17] Instead of using lunar orbit to accomplish these scientific objectives, we proposed a Lagrange point based space observatory to take advantage of its unique location that could provide a nearly unobstructed view of the lunar surface as well as earth's upper atmosphere and plasmasphere. Spacecraft at this position will have access to sunlight to generate power all the time without getting eclipses. As spacecraft will not face any eclipses it will help to avoid the worst-case scenario of the thermal environment (alternate hot and cold weather) of the spacecraft. However, an efficient active or passive thermal control system must be mounted on spacecraft to maintain the performance of electronic equipment which are design to operate within a certain temperature range.

### 3. <u>Stability Analysis and System Model</u>

Lagrange Points are regions in space where the gravitational force of two large masses precisely equals the centripetal force required for a small object to move with them. These points in space can be used by spacecraft to optimize the fuel consumption needed to remain in orbit. There are five special points where a small satellite can orbit in a constant pattern. Of the five Lagrange points, three are unstable points labeled as L1, L2, and L3 and two are stable points labeled L4 and L5. A satellite at L1, L2, or L3 is meta-stable, like a ball sitting on top of a hill. A tiny push may force the spacecraft to orbit around a metastable Lagrange point, which is called halo orbit. The L1 and L2 points are unstable on a time scale of approximately 23 days, which requires satellites orbiting these positions to undergo the regular course and attitude corrections. [26]

The dynamical systems of Spacecraft's three-body problem contain quasi-periodic orbits following the trajectories of the Lissajous orbits in L1 Earth-Moon Lagrange points with large amplitude Lissajous orbits decreasing the solar interference in communications for the space weather observatory. It also keeps the observatory out of the Earth's shadow and therefore provides continuous lighting for its solar panels. [26] [29] A linear stability analysis around Lagrange point L1 is required. It is important to resolve small deviations from equilibrium and to linearize the equation of motion around each equilibrium solution. The disturbance of the Earth-Moon system varies by bending the object's trajectory into a stable orbit around Lagrange's point L1. The Spacecraft at Lagrange point L1 requires a posture to maintain its position. Although the Lagrange point L1 is unstable, there are halo orbits around this point in a three-body system. Stability is achieved when the Coriolis acceleration curves the path into a path around the Lagrange point L1. [29] [31] Course corrections are required for the Earth-Moon system, to ensure the station-keeping of the spacecraft.

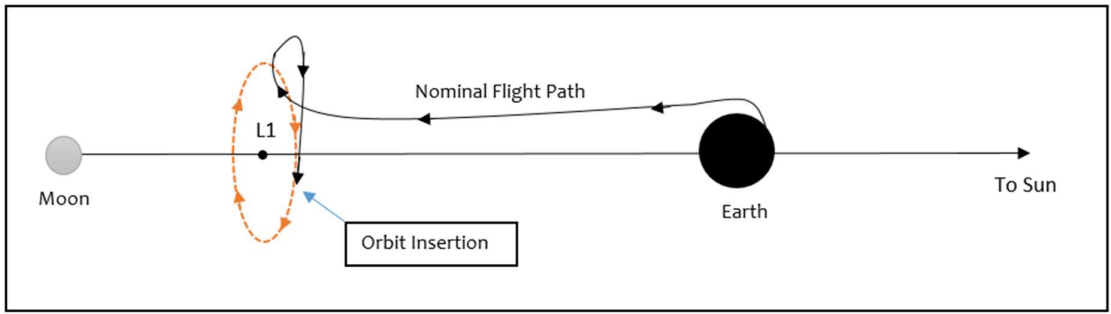

**Figure 2. Transfer Trajectory to Earth-Monn Lagrange Point L1.**

Some precise and continuous stationkeeping (SK) methods must be employed due to the unpredictable nature of the Lissajous orbit. Additionally, Certain missions located around the Earth-Moon Lagrange points require the reliability of such SK maneuvers to be extremely high. Stationkeeping methods are categorized into two major types, the tight control method refers to the technique that alters more than two parameters or Δv components to bring spacecraft back to a three-dimensional nominal path. The second is the "loose" control technique that is based on the "orbital energy balancing" strategy. The two control techniques differ only in the number of Δv components that are varied. The tight control technique varies two or more than two Δv components while the loose control technique varies only one to achieve a nominal path of spacecraft. A loose control strategy with orbital energy balancing used in SOHO(Solar and Heliospheric Observatory) is only concerned about one critical parameter to remove the unstable part of the orbit.[27] If the orbital energy of spacecraft is too high then the distance between spacecraft and the desired liberation point orbit will increase escaping towards the sun. If the orbital energy of spacecraft is too small then spacecraft will fall back towards the earth from orbit. [28] SK maneuvers are needed to correct for orientation requirements, orbit determination errors, Attitude Control Errors, uncertainty in perturbation forces, forces of thrust, and third body forces. In the Earth-Moon system, The uncertainty includes solar radiation pressure, the forces induces by magnetic fields, ocean tides, the earth's albedo effect, and possible errors in the stationkeeping maneuver execution. [30] The cost of these SK maneuvers can be reduced by performing frequent tiny maneuvers every week or two.

## 4. Simulation Results

A detailed analysis is needed to see the performance of spacecraft in the Circular Restricted Three-Body Problem (CR3BP). The Mission scenario is simulated using System Tool Kit (STK Astrogator) with appropriate implementation of the CR3BP force model. STK astrogator helps to study modeled environments and contributes to critical systems of mission design. It also supports the early concept design, selection as well as operation. [1] Spacecraft trajectories can be easily designed and converged by using different segments of the Mission Control Sequence (MCS)Tool of STK astrogator. Simulation allows visualizing the actual digital mission environment as shown in figure 3. Based on the data obtained by simulation, necessary attitude determination techniques could be implemented. This data could also be used for antenna pointing and solar panels orientation.

Here, we have simulated the whole scenario for the moon observatory from launch to the halo orbit insertion and up to four revolutions around the langrage point L1. The launch takes place from SDSC(Satish Dhawan space center, India) after 41.317 minutes(2,479 seconds) Spacecraft performed Trans-Lunar Injection(TLI) which put trajectory of spacecraft near Lagrange point L1. The relevant mission control segment data of maneuver summary is as shown in Table 1.

| Maneuver No. | Maneuver name | Start time (UTC) | Propulsion type (Engine Model) | Finite Burn Duration (Seconds) | Delta-V (m/sec) |
|---|---|---|---|---|---|
| 1 | Launch | 27 Sep 2021   03:28:42.339 | High Thrust and Low Isp | 989 | - |
| 2 | Tras-Lunar Injection (TLI) | 27 Sep 2021   04:13:38.471 | Constant Thrust and Isp | 56529.97 | 3120.21 |
| 3 | Lissajous Orbit Insertion | 01 Oct 2021   14:45:28.726 | Constant Thrust and Isp | 17661.04 | 672.13 |
| 4 | Impulsive Maneuver (SK) | 13 Oct 2021   00:21:24.631 | Constant Thrust and Isp | 78.665 | 2.67 |

**Table 1. Mission Control Sequence Summary.**

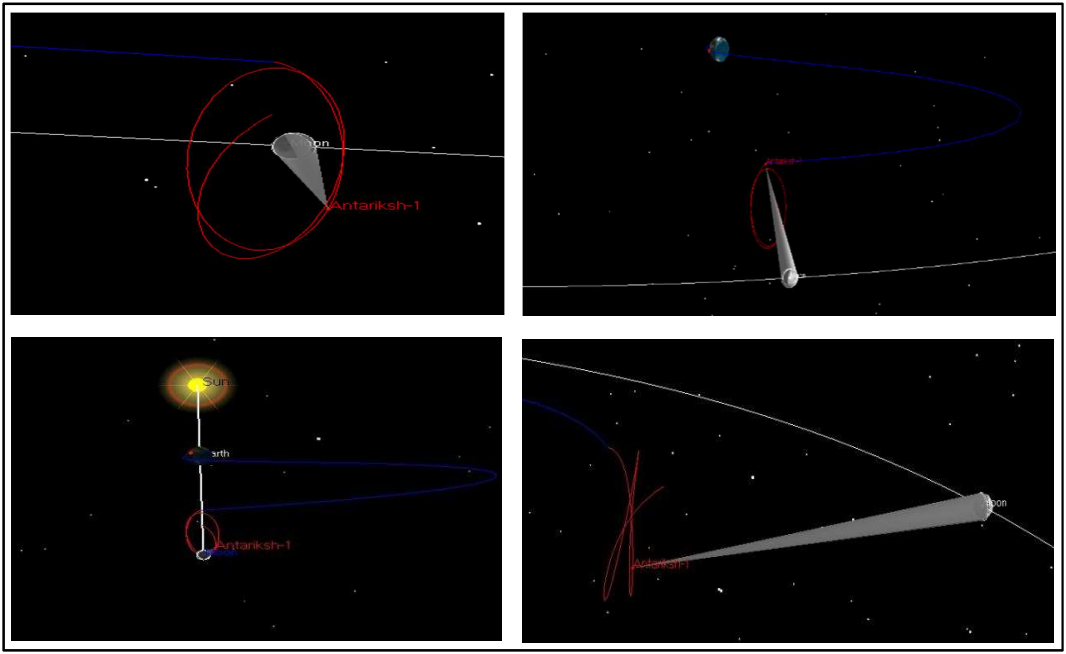

**Figure 3. Digital Mission Analysis of spacecraft in Lissajous orbit.** A space weather observatory is placed in L1 Lagrange point, to monitor Earth's magnetotail effects on the Moon.

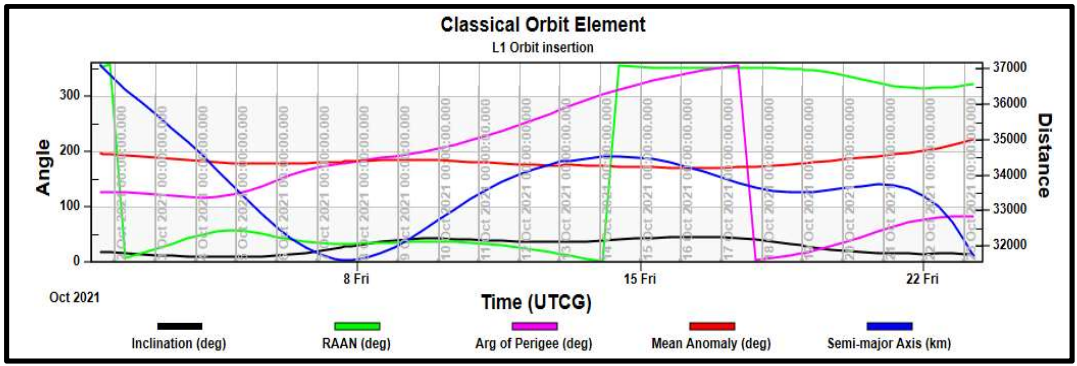

**Figure 4. Classical Orbital Elements of the spacecraft after Orbit insertion.**

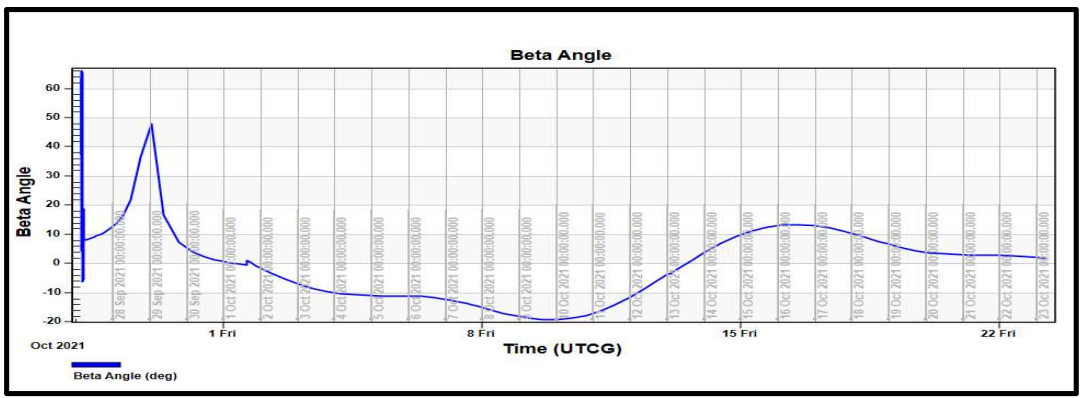

**Figure 5. Beta angle result from launch to Lissajous orbit insertion segment.**

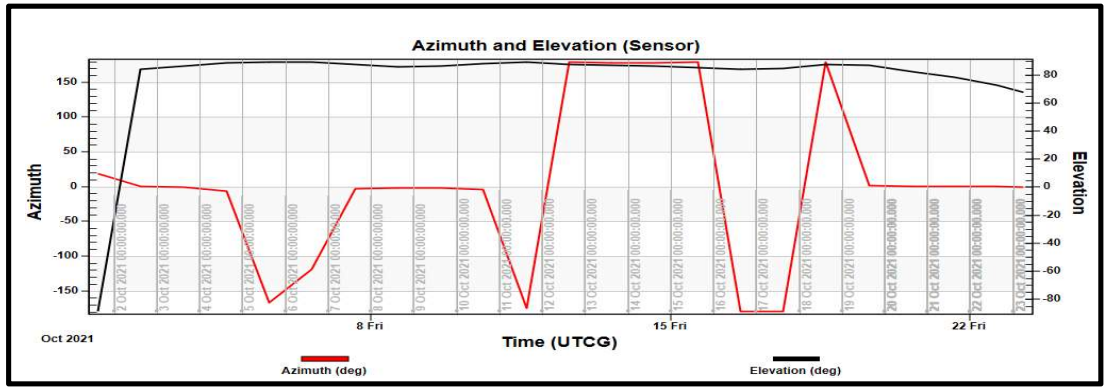

Figure 4. Remote Sensing Performance of Space weather observatory at Earth-Moon L1 point.

## 5. Conclusion

The Spacecraft placed in L1 Lagrange point provides more accurate weather observations of the moon surface. An advantage of using a Lagrange point observation station is that spacecraft near this point get an almost unobstructed view of the lunar surface and earth's atmosphere and plasmasphere. Numerical methods are integrated with simulation to analyze and optimize the three-body equilibrium between the Spacecraft, Earth, and the Moon. However, the trajectory optimization and the linearization of the equation of motion depends on the velocity and the Lagrange point stability analysis. The orbit around the Lagrange points is unstable which persuades the requirement of stationkeeping maneuver to maintain the position of a spacecraft. In other words, the spacecraft placed at Lagrange point will fall out of orbit unless course corrections are made. Although, effective stationkeeping techniques could be implemented on satellite attitude control systems considering both Tight control and Loose control techniques have their particular advantages and disadvantages. However, based on the delta-V expenditure i.e. Δv required (m/sec per year), maneuver cost in terms of fuel used, and frequent tiny course correction burns, an efficient control system could be designated to increase mission life. The spectroscopic data of geological, topological, and mineralogical surveillance of the moon received from the space weather observatory will allow astronomers to study further the Moon rusting, its consequences on the Earth's environment, and allow other planetary missions to carry out more advanced astronomical observations.